# Loop Bifurcation and Magnetization Rotation in Exchange Biased Ni/FeF$_2$

Justin Olamit[1], Elke Arenholz[2], Zhi-Pan Li[3], Oleg Petracic[3,4], Igor V. Roshchin[3],

R. Morales[3,5], Xavier Batlle,[3,6] Ivan K. Schuller[3], and Kai Liu[1,*]

[1]*Physics Department, University of California, Davis, CA 95616*
[2]*Advanced Light Source, Lawrence Berkeley National Laboratory, Berkeley, CA 94720*
[3]*Physics Department, University of California- San Diego, La Jolla, CA 92093*
[4]*Angewandte Physik, University Duisburg-Essen, 47048 Duisburg, Germany*
[5]*Departamento de Fisica, Universidad de Oviedo, Oviedo 33007, Spain*
[6]*Departament de Física Fonamental, Universitat de Barcelona, 08028 Barcelona, Catalonia, Spain*

Abstract

Exchange biased Ni/ FeF$_2$ films have been investigated using vector coil vibrating sample magnetometry as a function of the cooling field strength $H_{FC}$. In films with epitaxial FeF$_2$, a loop bifurcation develops with increasing $H_{FC}$ as it divides into two sub-loops shifted oppositely from zero field by the same amount. The positively biased sub-loop grows in size with $H_{FC}$ until only a single positively shifted loop is found. Throughout this process, the negative/positive (sub)loop shift has maintained the same *discrete* value. This is in sharp contrast to films with twinned FeF$_2$ where the exchange field *gradually* changes with increasing $H_{FC}$. The transverse magnetization shows clear correlations with the longitudinal sub-loops. Interestingly, over 85% of the Ni reverses its magnetization by rotation, either in one step or through two successive rotations. These results are due to the single crystal nature of the antiferromagnetic FeF$_2$, which breaks down into two opposite regions of large domains.

Exchange bias in ferromagnet / antiferromagnet (FM/AF) bilayers has attracted intense interest due to its elusive mechanism and important applications in "spin-valve" type devices [1-3]. It is generally believed that magnetic domains, in the AF and FM layers, are crucial [4]. One manifestation of the importance of the FM domains is the loop bifurcation seen in zero field cooled (ZFC) FM/AF bilayers where both positively and negatively biased sub-loops coexist [5-7]. In those studies, the opposite FM domains at remanence were frozen in upon cooling, resulting in oppositely biased regions in the film.

Key length scales in exchange biased systems are the domain size in both FM and AF layers, $D_{FM}$ and $D_{AF}$. For example, it has been shown in FM/AF nanostructures where the FM domain size is reduced through nanopatterning that exchange bias can be significantly altered [8]. In contrast, there have been limited experimental studies on the AF domain size dependence of exchange bias [9], largely due to the difficulty in characterizing and manipulating AF domains in FM/AF systems [10-12]. The domain sizes may also affect the shape of the hysteresis loop. Recent experiments on model systems using synthetic antiferromagnets have shown that larger domain sizes could lead to a loop bifurcation [13,14].

In this work, we demonstrate that a bifurcated hysteresis loop can be realized in field cooled (FC) Ni/epitaxial-FeF$_2$ bilayers. In comparison, the bifurcation is absent in FC Ni/twinned-FeF$_2$ and Ni/polycrystalline-FeF$_2$, which are expected to have smaller AF domain sizes. Unlike the aforementioned bifurcation seen in ZFC bilayers caused by opposite FM domains [5-7], here the FM is a single domain and the bifurcation is due to the opposite AF domains frozen in during cooling and consequently can be tuned by the cooling field. Additionally, the Ni layer reverses its magnetization mostly via rotation.

For this study, we have prepared thin films of Al (76 Å) / Ni (210 Å) / $FeF_2$ (500 Å) simultaneously onto single crystal $MgF_2$ (110), MgO (100) and Si (100) substrates, to respectively achieve untwinned epitaxial (110), twinned (110), and polycrystalline $FeF_2$ [15-18]. All samples have been grown by electron beam evaporation, using conditions similar to those reported in earlier publications [16,18]. The $FeF_2$ layer was deposited at 200 - 300 ºC while the Ni and Al layers were grown at 150 ºC. The crystal structures of $FeF_2$ and Ni (always polycrystalline) have been confirmed by x-ray diffraction. The $FeF_2$ (110) rocking curve full-width at half maximum (FWHM) is about 3.9º (using Cu K$\alpha$) for twinned $FeF_2$ and < 1.2º for untwinned epitaxial $FeF_2$. This is consistent with x-ray reflectivity measurements by Shi and Lederman where the in-plane coherence length in similarly prepared twinned $FeF_2$ was determined to be about 60-100 Å and that in untwinned $FeF_2$ was ~ 280 Å [17].

Field cooling and magnetic measurements have been performed in a vibrating sample magnetometer. The samples were cooled from 150 K (above 78K, the $FeF_2$ Néel temperature) to 15 K in different cooling fields $H_{FC}$ (2 –15 kOe). The cooling and measurement field direction is always the same and in the film plane. Both longitudinal moment ($m_{//}$, the component parallel to $H$) and transverse moment ($m_{\perp}$, the component in the film plane but perpendicular to $H$) have been measured with vector detection coils. At 15 K, the polycrystalline sample has an exchange field $H_E$ of about -300 Oe, with no appreciable dependence on $H_{FC}$. The twinned film displays a larger $H_E$ of -700 Oe in $H_{FC}$ =1 kOe and -600 Oe in $H_{FC}$ =15 kOe (measured with the field along MgO [001]). Below we concentrate on the Ni/epitaxial-$FeF_2$, which is oriented with the field along the $FeF_2$ spin axis, [001] direction.

The magnetic hysteresis loops of the longitudinal (solid circles) and transverse (open circles) moments under different $H_{FC}$ are shown in Fig. 1. At $H_{FC}$ = 2 kOe, a single longitudinal loop is found, negatively biased with a $H_E \approx$ -1 kOe (Fig. 1a) which is larger than that in Ni/twinned-FeF$_2$. As $H_{FC}$ is increased (2 kOe < $H_{FC}$ < 15 kOe), a loop bifurcation develops as the loop splits into two oppositely shifted sub-loops separated by a plateau in the middle. The two sub-loops have similar coercivity and symmetric exchange fields, $H_E \approx \pm 1$ kOe (Fig. 1b & 1c). Note that as $H_{FC}$ increases, the sub-loops maintain their discrete exchange fields, at ±1 kOe, only their relative sizes change. For example, the negatively biased sub-loop accounts for ~ 85% of the total Ni magnetization at $H_{FC}$ = 5 kOe, but only ~ 35% at $H_{FC}$ = 7.5 kOe. Finally at $H_{FC}$ = 15 kOe, the loop is completely positively shifted with $H_E \approx$ +1 kOe. Thus an increasing $H_{FC}$ drives the sample from negative to positive bias while the middle plateau moves in a "top-to-bottom" fashion. This is in contrast to Ni/twinned-FeF$_2$ and Fe/twinned-FeF$_2$ [15] samples where $H_{FC}$ drives the whole loop continuously (for certain samples from negative to positive bias) in a "left-to-right" fashion.

The transverse loops show peaks at the fields corresponding to the switching fields of the longitudinal (sub)loops. We will use "upward" and "downward" to represent directions in the film plane that are perpendicular to the applied field, corresponding to positive and negative transverse moment, respectively. For $H_{FC}$ = 2 kOe, the transverse loop shows two positive peaks at $H \approx$ -1 kOe – one for each branch of the field cycle (Fig. 1a). Both peaks have large magnitudes, ~ 82% of the saturation moment, $m_s$. This indicates that most of the sample reverses its magnetization via rotation, as opposed to domain nucleation and motion. In films with twinned and polycrystalline FeF$_2$, we usually observe a smaller transverse peak, ~ 10-60% of the total magnetization. As $H_{FC}$ is increased, a pair of peaks at $H \approx \pm 1$ kOe is observed for *each* field-sweep direction (Fig. 1b & 1c). For the decreasing-field branch, the transverse moments point downward while those in the increasing-field branch point upward. Note that once the positively biased sub-loop appears, magnetization reversal from positive saturation changes from upward (Fig. 1a) to downward rotation (Figs. 1b-1d). For different $H_{FC}$, the transverse peak locations stay constant; only the peak magnitudes change. Finally, for $H_{FC}$ = 15 kOe, the transverse hysteresis loop shows two peaks at $H \approx +1$ kOe – again one for each field sweep branch (Fig. 1d). The moments always point downward (negative), whose maximum magnitude is also ~ 82% $m_s$.

The behavior of the longitudinal (sub)loops and transverse peaks can be attributed to the existence of two regions of FM domains, FM-A & FM-B, which respectively correspond to the negatively and positively biased sub-loop. The sizes, or fractions, of each region are apparent in both the longitudinal and transverse loops. In the longitudinal direction, the domain fraction can be defined as

$$m_{//}^A \text{ fraction} = |m_{//}^A|/2m_s \qquad (1)$$

$$m_{//}^B \text{ fraction} = |m_{//}^B|/2m_s \qquad (2)$$

where $m_{//}^A$ and $m_{//}^B$ are the moments associated with the sub-loops, as schematically shown in Fig. 2a. In the transverse direction, the domain fraction can be defined as

$$m_{\perp}^A \text{ fraction} = |m_{\perp}^A|/m_s \qquad (3)$$

$$m_{\perp}^B \text{ fraction} = |m_{\perp}^B|/m_s \qquad (4)$$

where $m_\perp^A$ and $m_\perp^B$ are the local maximum (peak) moment value near $H = \pm 1$ kOe, as schematically shown in Fig. 2b [19].

As shown in Fig. 2c, there is a clear one-to-one correlation between the domain fractions in longitudinal and transverse loops. At $H_{FC} = 2$ kOe, there is no kink in the longitudinal loop, thus the $m_{//}^A$ fraction is 1.00, and the corresponding $m_\perp^A$ fraction is 0.82. At $H_{FC} = 15$ kOe, similar $m_{//}^B$ and $m_\perp^B$ fractions are observed. Between 2 and 15 kOe, with increasing $H_{FC}$, the FM-B fraction gradually increases at the expense of FM-A. This is manifested in the longitudinal loop as the middle plateau sweeps downward, and in the transverse loop as the negative $m_\perp$ peak at $H \sim +1$ kOe grows in size. It is interesting to note that $|m_\perp^A| + |m_\perp^B|$ is ~85% $m_s$, as shown in Fig. 2c inset. The correlation demonstrates that at any $H_{FC}$, ~85% of the FM reversal is by rotation, either in one step or by two successive rotations!

The two FM domain regions at intermediate $H_{FC}$ are coupled to the underlying AF domains, forming two regions of FM/AF domains with opposite unidirectional anisotropies, as schematically illustrated in Fig. 3. Within each region, unidirectional exchange anisotropy is induced by field cooling along the AF easy axis, confirmed by in-plane angular dependence of the exchange field. As shown in Fig. 3a, the applied field is pointing to the right, defined as the positive direction for longitudinal moment $m_{//}$; 90º counterclockwise is defined as the positive (or upward) direction for the transverse moment $m_\perp$. Also shown is an inevitable but exaggerated misalignment between the anisotropy directions and the applied field (dashed lines) during cooling and measurement. The opposite unidirectional anisotropies are results of the competition between the antiparallel interfacial coupling between Ni and $FeF_2$ and the Zeeman energy that aligns the interfacial AF spins [15, 16]. During field cooling with a small $H_{FC}$ = 2 kOe, the aligned FM spins couple antiparallel to the interfacial AF spins (domain A). Once frozen below the AF Néel temperature, this leads to a hysteresis loop shifted to the left, giving rise to a negative exchange bias. When cooling with a large $H_{FC}$ = 15 kOe, the interfacial AF spins are forced to align with $H_{FC}$ (domain B). Once frozen in, these spins then try to align the FM spins opposite of the external field through their interfacial coupling, leading to a positive exchange bias. At 2 kOe < $H_{FC}$ < 15 kOe, these two types of domains coexist, resulting in a superposition of negatively and positively biased sub-loops with *discrete* exchange fields.

The domain structures are also responsible for the rotation behavior of the FM spins found in the transverse loops. At $H_{FC}$ = 2 kOe, domain A is favored. Its unidirectional exchange anisotropy $H_E^A$ has a small upward component due to the aforementioned misalignment. This component then guides the magnetization to rotate upward during reversal from +$H$ to $H_E^A$ and –$H$, as illustrated in Fig. 3b, and then –$H$ to $H_E^A$ and +$H$ during the returning path, leading to the positive $m_\perp$ peaks at $H \sim -1$ kOe. Since $H_E^A$ is in close alignment with +$H$, as opposed to –$H$, the reversal from/to positive saturation is always sharper than that from/to negative saturation, as seen in Fig. 1a. Similarly, at $H_{FC}$ = 15 kOe, domain B is favored. Its unidirectional anisotropy $H_E^B$ has a downward component, forcing downward magnetization rotation from +$H$ to $H_E^B$ and –$H$ (Fig. 3c) and resulting in the negative peak at $H \approx +1$ kOe in the transverse loop.

At $H_{FC}$ = 5 and 7.5 kOe, both types of domains are present in the AF layer. At large applied fields, the FM layer is single domained. As the field is reduced, the AF domain structures break the FM layer into domains where the FM reversal is dictated by the first FM domain to rotate in each field sweep direction. Along the decreasing-field sweep, the positively shifted sub-loop associated with domain B experiences reversal first. The downward rotation leads to a negative $m_\perp$ peak at $H \approx +1$ kOe. As $H$ continues to decrease, the FM-A spins start to rotate. The small upward component of $H_E^A$ alone would pull the moments to rotate upward to form a positive $m_\perp$ peak at $H \approx -1$ kOe. However, the already rotated FM-B now drag the FM-A spins to rotate in the same orientation, so as not to create a domain wall that is over 180 degrees. Eventually both domains are saturated along $-H$. In the increasing-field sweep, FM-A spins are the first to reverse at -1 kOe. Now the small upward component of $H_E^A$ indeed pulls the FM spins to rotate upwards, which in turn drag FM-B spins to also rotate upward (rather than downward) at + 1 kOe. Hence, the $m_\perp$ peaks in general are both negative and both positive for the decreasing- and increasing-field sweep, respectively [20].

There are two important length scales which control the behavior of exchange biased FM/AF layers. These are the FM and AF domain size, $D_{FM}$ and $D_{AF}$. For the case where $D_{FM} \gg D_{AF}$ during switching, each FM domain couples to many AF domains, therefore it is only sensitive to the averaged AF interface spins. With increasing $H_{FC}$, the shift in the major loop is always the *averaged* exchange bias over the length scale of $D_{FM}$, thus it changes gradually. This is likely the case in previous studies on twinned $FeF_2$, where the small twin sizes and the orthogonal twinning encourage the formation of small AF domains with orthogonal anisotropy directions [15, 16].

In the opposite limit, i.e., where $D_{FM} \leq D_{AF}$ during switching, each FM domain couples to a large, uniform collection of interfacial AF spins so that the total influence (and consequently the exchange bias) on the FM domain is greater than that for the $D_{FM}$ $>> D_{AF}$ case. Furthermore, when opposite AF domains are present, each FM domain still couples to a uniform AF structure and is discretely biased. The sum of the FM domains results in sub-loops. This is consistent with the findings in synthetic antiferromagnets where it has been shown that a loop bifurcation develops as the domain size becomes larger [13, 14]. For the Ni/epitaxial-FeF$_2$ bilayers studied here, the larger in-plane coherence length and the lack of orthogonal twinning encourage the formation of large domains, leading to a larger $D_{AF}$ than that in twinned-FeF$_2$. Although the actual $D_{AF}$ here is yet to be determined, large AF domains (up to ~mm scale) have been seen in bulk single crystal fluorides [21, 22]. Additionally, the large AF domains also encourage the FM domains to remain intact during magnetization reversals. This explains the large domain fractions (~85%) seen in the transverse hysteresis loops that reverse by rotation. Indeed, recent magneto-optical Kerr effect studies have shown the existence of ~mm scale domain structures in similar samples [18], and simulations have reproduced the loop bifurcation for systems with large AF domains [23].

It is interesting to note that the exchange field is larger in Ni/untwinned-FeF$_2$ than in Ni/twinned-FeF$_2$, where the latter is expected to have a smaller AF domain size. This is opposite to some of the theoretical predictions [4] as well as experimental results in Co/LaFeO$_3$ [9]. The key issue might be whether the exchange field depends most critically on the AF domain size or the interfacial spin density. This is the subject of an ongoing study.

In summary, we have observed large and discrete values of exchange field in exchange coupled Ni / epitaxial-FeF$_2$ films. With increasing cooling field, the sample divides into two regions of oppositely biased domains. The positively-biased region grows at the expense of the negatively biased one, eventually taking over the entire sample. The magnetization reversal is predominantly by rotation. These results are attributed to a large $D_{AF}$, compared to $D_{FM}$, that has suppressed the usual averaging effect observed in twinned samples.

Work at UCD was supported by University of California CLE and NEAT–IGERT fellowship (J.O.). The acquisition of a vibrating sample magnetometer which was used extensively in this investigation was supported by NSF Grant No. EAR-0216346. Work at LBNL and UCSD were supported by DOE. Financial supports from Alexander-von-Humboldt Foundation (O. P.), Cal-(IT)$^2$ (Z.-P. L.), Spanish MECD (R. M., X. B.), Fulbright Commission (R. M.), and Catalan DURSI (X. B.) are acknowledged.

**Figure Caption**

Fig. 1. Longitudinal ($m_{//}$, solid circles) and transverse ($m_{\perp}$, open circles) magnetic hysteresis loops of Ni/FeF$_2$ field cooled in $H_{FC}$ of (a) 2 kOe, (b) 5 kOe, (c) 7.5 kOe, and (d) 15 kOe. The moments are normalized to the saturation moment $m_s$. Arrows indicate the field-cycle sequence for the transverse loop.

Fig. 2. Illustrations of the determination of (a) longitudinal and (b) transverse domain fractions, as defined in Eqn. 1-4 in the text. The asymmetric transverse peaks are caused by a misalignment between the magnetic field and the anisotropy axis, as explained in the text. (c) Correlation between the longitudinal and transverse domain fractions, for all cooling fields. Solid and open symbols represent domain A and B, respectively. Solid line is a guide to the eye with a slope of 0.85. The inset of (c) shows the sum of $m_{\perp}^{A}$ and $m_{\perp}^{B}$ fractions, where the horizontal line indicates about 85% of the FM reverse by rotation.

Fig. 3. (a) Schematic of the two regions of large FM/AF domain structures, with respect to the applied field $H$ direction. A misalignment between the anisotropies and the field direction, illustrated by the dashed lines, is exaggerated. The orientation of the easy direction in (b) domain A and (c) domain B differentiates the magnetization rotation direction.

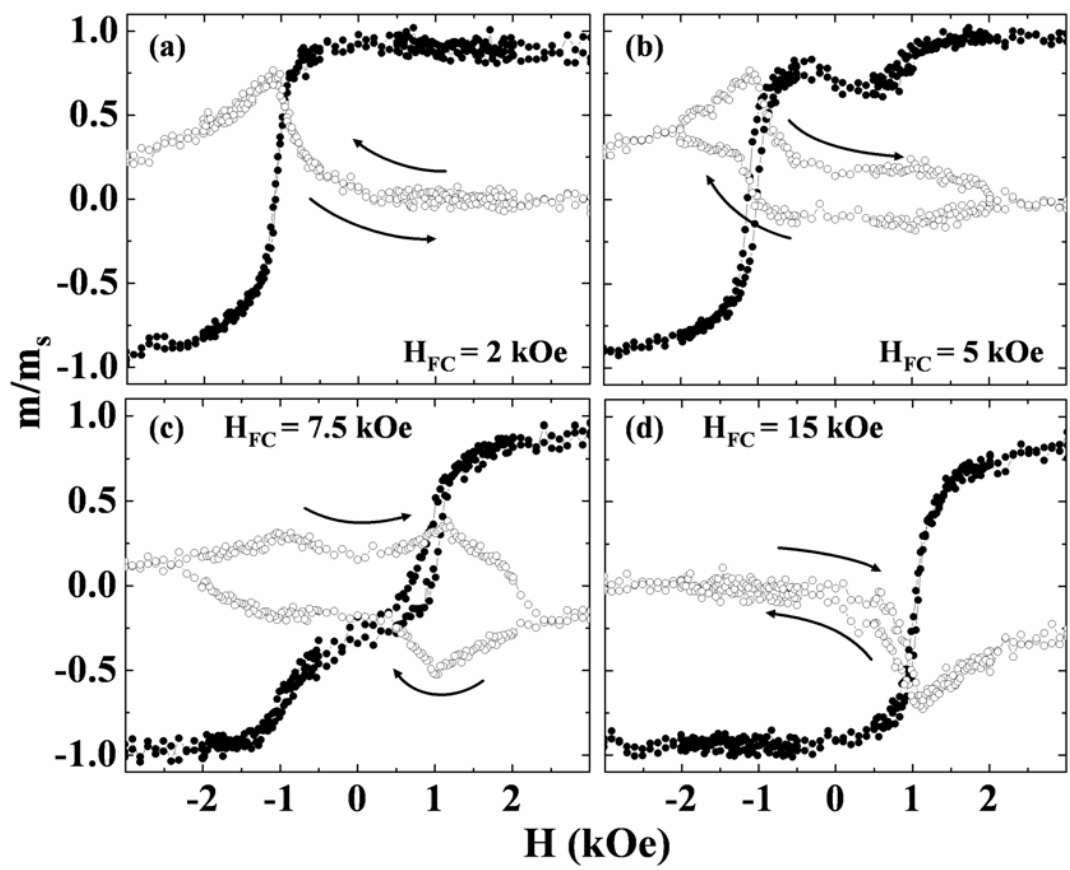

**Fig. 1**

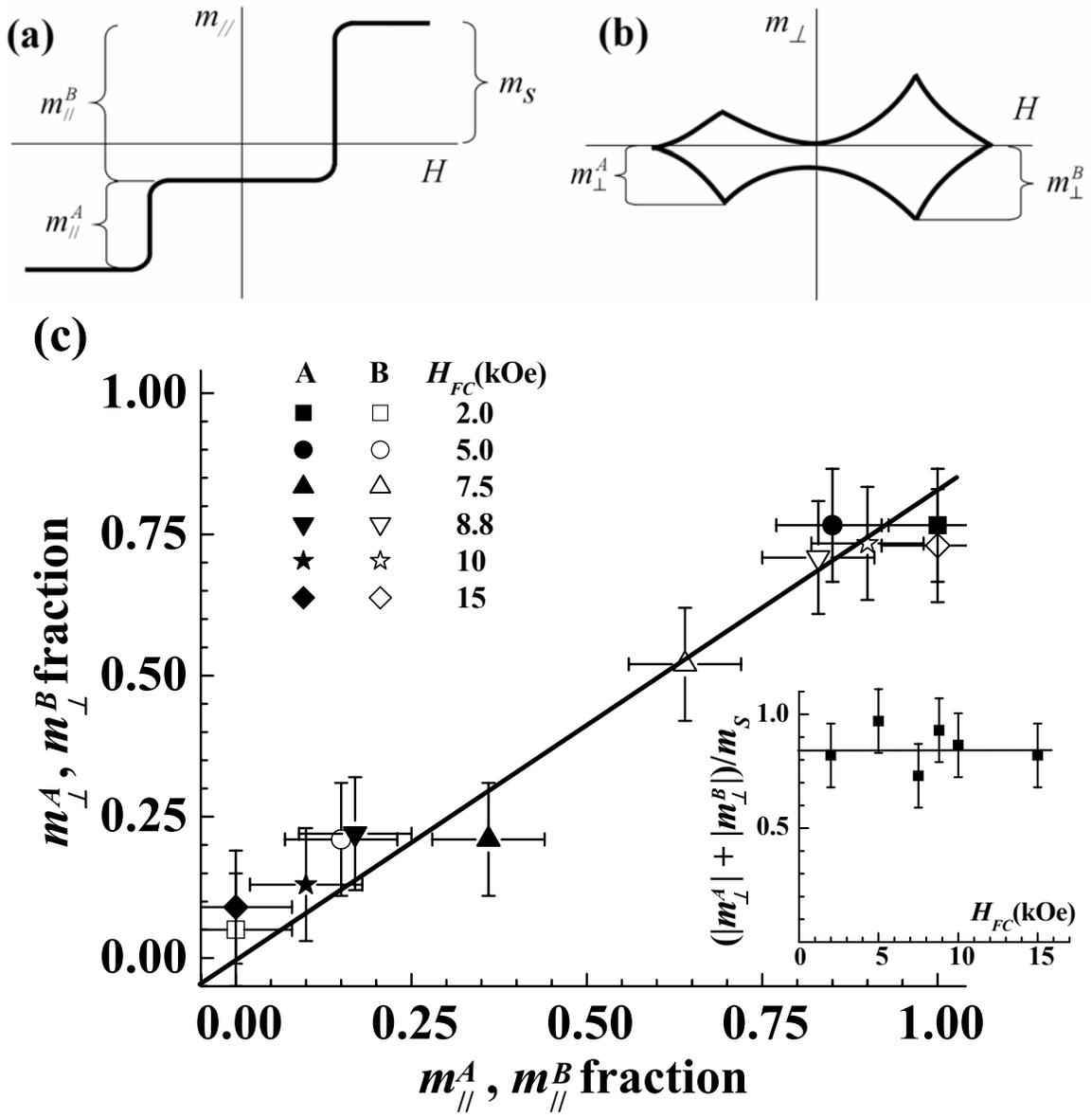

Fig. 2

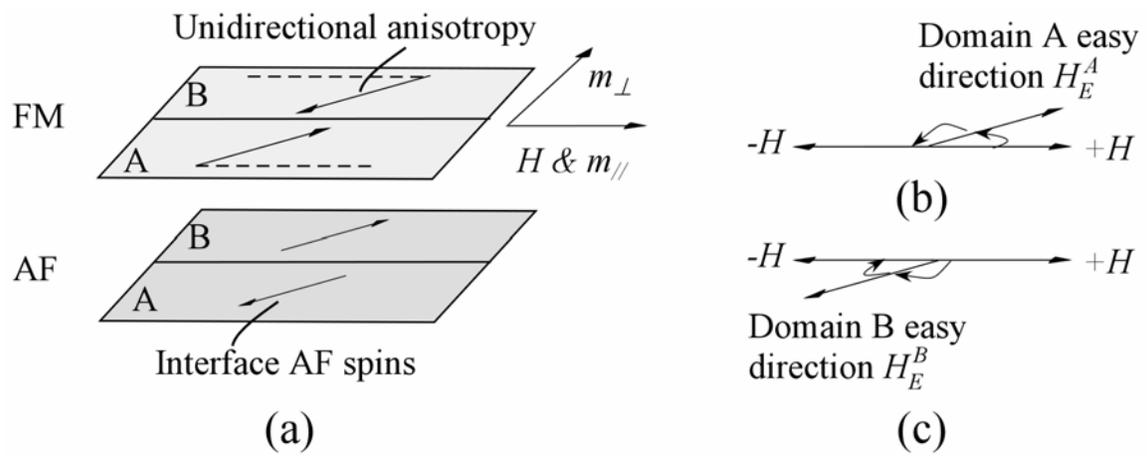

**Fig. 3**